\documentclass{aa}  
\usepackage{graphicx}
\usepackage[colorlinks,allcolors=blue]{hyperref}
\bibpunct{(}{)}{;}{a}{}{,}
\usepackage{bm,natbib,url,wasysym}
\usepackage[varg]{txfonts}
\usepackage{xcolor}
\usepackage{subcaption}
\usepackage{mathtools}

\usepackage{graphicx}
\usepackage{txfonts}

\usepackage[normalem]{ulem}

\newcommand{\nssa}{\nu'_{\rm{ssa}}}
\newcommand{\Leo}{L_{\rm{e,0}}^{\rm 'inj}}
\begin{document}

\title{An expanding one-zone model for studying blazars emission}

%\subtitle{From radio to $\gamma$-ray}

\author{S.Boula
\inst{1}\fnmsep\thanks{ \email{stboula@ifj.edu.pl}. Current address: Institute of Nuclear Physics Polish Academy of Sciences, PL-31342 Krak{\' o}w, Poland
}
\and
A. Mastichiadis\inst{1}
}

\institute{
Department of Physics, National and Kapodistrian University of Athens, GR 15783 Zografos, Greece}
\date{Received September XX, XXXX; accepted March XX, XXXX}

% \abstract{}{}{}{}{} 
% 5 {} token are mandatory

\abstract
{Blazars, a sub-category of Active Galactic Nuclei, are characterized by their non-thermal variable emission. 
This emission extends over the whole electromagnetic spectrum and is a consequence of particle acceleration inside their relativistic jets. However, especially the relation of radio emission to that at higher frequencies remains an open question. }
{Observations of blazar emission show that the location of radio might be very different from the one where the rest of the spectrum is produced and often requires separate modeling. 
We aim to produce both emissions within the context of one model.}
{We construct a self-consistent one-zone expanding leptonic model for studying the connection between the radio emission and the emission at higher frequencies and we apply it to the flaring states of blazars. Assuming an accelerating episode as the source moves down the jet and expands, we numerically study the electron evolution as they lose energy due to adiabatic expansion and synchrotron/inverse Compton radiation.}
{We find that high-frequency radiation mimics the electron injection and is mainly produced close to the acceleration site where cooling is strong. In contrast, the radio emission is produced further down the jet when the emitting region has become optically thin to synchrotron self-absorption due to expansion. We present briefly the role of the initial parameters, such as the magnetic field strength, the electron luminosity and expansion velocity, on the localization of the radio emission site. We show that the expanding one-zone model is inherently different from the non-expanding one and, in addition, it requires more parameters. For example, we apply our approach to the observational data of a Mrk 421  $\gamma$-ray - radio flare observed in 2013.
}
{}


\maketitle
\section{Introduction}\label{s1}
Blazars, the most extreme subclass of Active Galactic Nuclei (AGN), have their relativistic jets pointing towards the observer. Blazar emission is non-thermal, it spreads over the entire electromagnetic spectrum and it is characterized by rapid variability, high optical polarization and apparent superluminal motion. One of the characteristic features of this emission is the shape of its Spectral Energy Distribution (SED), which usually has two components: a low-energy component extending from radio to UV/soft X-rays  and a high-energy one lying between hard X-rays and TeV $\gamma$-ray.

Modeling blazar SED implies the existence of a non-thermal population of particles. In many cases, blazar emission is explained by a relativistic electron population which accelerates and radiates inside a spherical region \citep[see the following review papers ][]{Pa2017,HLjets,Boettcher19,Blandford19, Mr20}. This is the so-called one-zone leptonic model and has been applied widely in modeling the blazar SED \citep[e.g.,][]{MG85,1996Inoue,MK97,KRM98,KM1999,Kusunose2000,Bottcher2002,2011tramacere,Asano2014,2016Finke, Boula19a}. In this approach the low-frequency component is related to synchrotron radiation while the higher energy one to inverse Compton scattering. Reproducing the observed SED leads to physical parameters constraints. 

The multi-wavelength monitoring programs are a unique way of probing the conditions in blazar jets and unveiling the physical processes responsible for the blazar non-thermal variable emission. High-quality data, e.g., time lags between radio and $\texorpdfstring{\gamma}{}$-ray emission, which are associated to significant correlations between these energy bands, contribute to the understanding of blazar anatomy \citep[e.g.,][]{P14, 2015Esposito,HP15, angelakis16, T18,2019Liodakis}. \\
\indent Despite this multi-year monitoring effort, there is still no consensus about the location of the high-energy (X-rays and $\gamma$-ray) emission in blazar jets. In particular, the rapid variability, which is an identifying property of blazars, suggests that the non-thermal emission is typically produced in regions of the jet with size $\ll$ 1 light day \cite[e.g.,][]{aharonian07, aleksic11}. Although electron synchrotron radiation produced in this region can explain the optical to X-rays part of the photon spectrum, in most of the cases it cannot account for the jet emission at low frequencies (e.g., $\nu \leq 10^{10}$ Hz); in such compact emitting regions, synchrotron radiation is typically self-absorbed \cite[e.g.,][]{VdL69,1969GS, TAM2001,2006Kaiser}. However, blazar jets are detected up to GHz frequencies having a power-law spectrum \citep[see also][]{2003tingay,Stawarz08,G17}. In order to solve this discrepancy, it has been proposed that the radio emission, in most cases, is not produced in the same region as the $\gamma$-ray. The radio-emitting area should be larger to be less opaque to synchrotron self-absorption \citep{M80,MG85, G85,potter1, M14,Potter18, Boula19b}. Still, the results of the various monitoring programs raise several questions about the origin of the radio emission of the jets, which in many cases is variable \citep{HB2001, KLC2010, GG2011, AT2012,2018Myserlisb}. 
\\
\indent Many theoretical studies are based on an idealized model of a conical steady radio jet, while the component which is associated with the variability of the source is related to shock waves that are propagating in the jet \citep{bk79,GM98b,GM98a,katarzynski03}. 
Assuming that electrons are accelerated in such shock fronts and are cooled as they move away from it, one gets that the highest frequency synchrotron component is emitted from a thin layer behind the shock. In contrast, the lower frequency component is produced in a larger volume, simply because the cooling time for higher-energy electrons is less than for the lower-energy ones, \cite{MG85}.
Furthermore, this scenario predicts time lags between high-frequency and low-frequency lightcurves and a specific evolution of the radio spectrum with time.
\cite{potter1,Potter18} proposed a model of a ballistic jet with a uniform structure to investigate the relationship between low and high-energy emission \citep[see also][]{katarzynski03,HBS15,RS16,2021zacharias}.
\\ 
\indent The goal of this work is to study the relationship between radio and $\gamma$-ray activity in blazars and to localize their emission site using a simple framework that allows a comprehensive search of the parameter space. More specifically, we use a time-dependent numerical code that was originally developed to treat the radiative transfer problem in a static spherical geometry \citep{MK95}, to compute the emission from an expanding blob of plasma, \citep[see also][]{ZW16, Boula19b}. This approach allows $\gamma$-ray to be produced more efficiently close to the electron acceleration site, whereas radio is produced much later when the blob has expanded enough to become optically thin to synchrotron self-absorption. The numerical code allows us also to study reaccelerating episodes and we show that depending on the distance, this can produce $\gamma$-ray flares, which might or might not accompanied by radio flares. 

The paper is structured as follows: in Section \ref{s2}, we present our new one-zone expanding model that considers all the leptonic physical processes. In Section \ref{s3}, we show the dependence on the initial parameters of the transition from the optically thick to the optically thin region due to synchrotron self-absorption. This method guides us to localize the onset of radio emission inside the jet. The initial values of the physical quantities of the source play a vital role in the transition from the optically thick to the optically thin region.  In Section \ref{s4}, we present a few characteristics examples for the flaring activity of blazars by assuming electron accelerating episodes and following their evolution. We also apply our model to the flaring episode of Mrk421 that occurred in 2013, explaining the time-lag between $\gamma$-ray and radio emission. In Section \ref{s5}, we discuss our results and finally, in Section \ref{s6}, we give the basic conclusions of our work. 

\section{One-zone expanding blob model}\label{s2}

We consider a blazar with a conical jet whose axis makes an angle $\theta_{\rm obs}$ to our line of sight. We assume that, due to some accelerating episode, relativistic electrons are injected in a spherical region, which we will refer to as "the blob", of initial radius $R'_{\rm 0}$.  

The blob moves along the jet axis with a velocity $u_{\Gamma}=\beta_\Gamma c\simeq c$ corresponding to a Lorentz factor $\Gamma = (1-\beta_{\rm \Gamma}^2)^{-1/2}$ and to a Doppler factor $\delta=[\Gamma (1-\beta_{\rm \Gamma} \cos \theta_{\rm obs})]^{-1}$. At the same time, it expands in a spherically symmetric way with velocity $u_{\rm exp}$ in its rest frame. Therefore, if we denote $t’$ the time as measured in the rest frame of the blob{\footnote {Primed symbols refer to the quantities as measured in the co-moving frame.}}, 
then following \cite{Boula19b} we can write for its instantaneous radius:
\begin{equation}
R’(t')=R’_{\rm 0} +u_{\rm exp}(t’-t’_{\rm 0})
\end{equation}
where $t’_{\rm 0}$ is the time of the blob formation. Similarly, the instantaneous distance $z=z'/\Gamma$, of the blob along the jet at the observer's frame will be given by
\begin{equation}\label{eq.zt}
z(t)=z_{\rm 0}+\beta_{\Gamma} c (t'-t'_{\rm 0})/\Gamma
\end{equation}
or 
\begin{equation}\label{eq.zR}
z(t)=z_{\rm 0}+\beta_{\Gamma} c [R'-R'_{\rm 0}]/(\Gamma u_{\rm exp})
\end{equation}
where  $z_{\rm 0}$ is the blob's initial distance from the central object.
Therefore both quantities, $R'$ and $z$, are functions of the time $t'$.

\begin{figure*}[!htbp]
\centering

\begin{subfigure}[b]{0.45\textwidth}
\centering
\includegraphics[width=10.cm,trim=60 4 4 6]{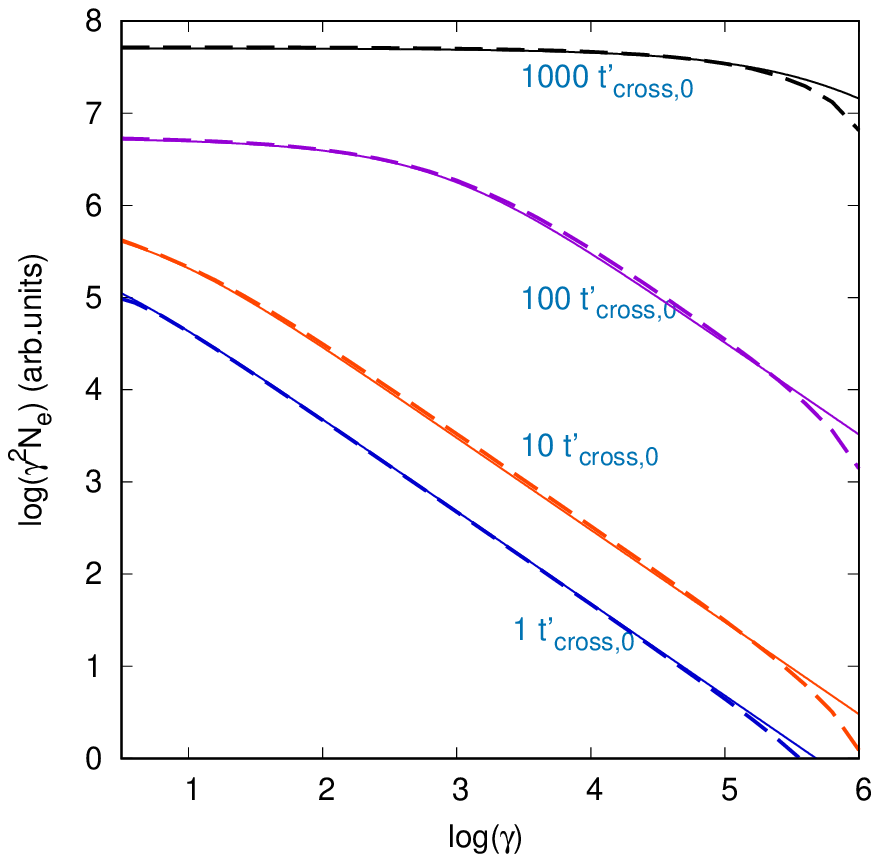}
\caption{}
\label{fig:comp-elec}
\end{subfigure}
\hspace{1cm}
\begin{subfigure}[b]{0.45\textwidth}
\centering
\includegraphics[width=10.5cm, trim=60 15 4 0]{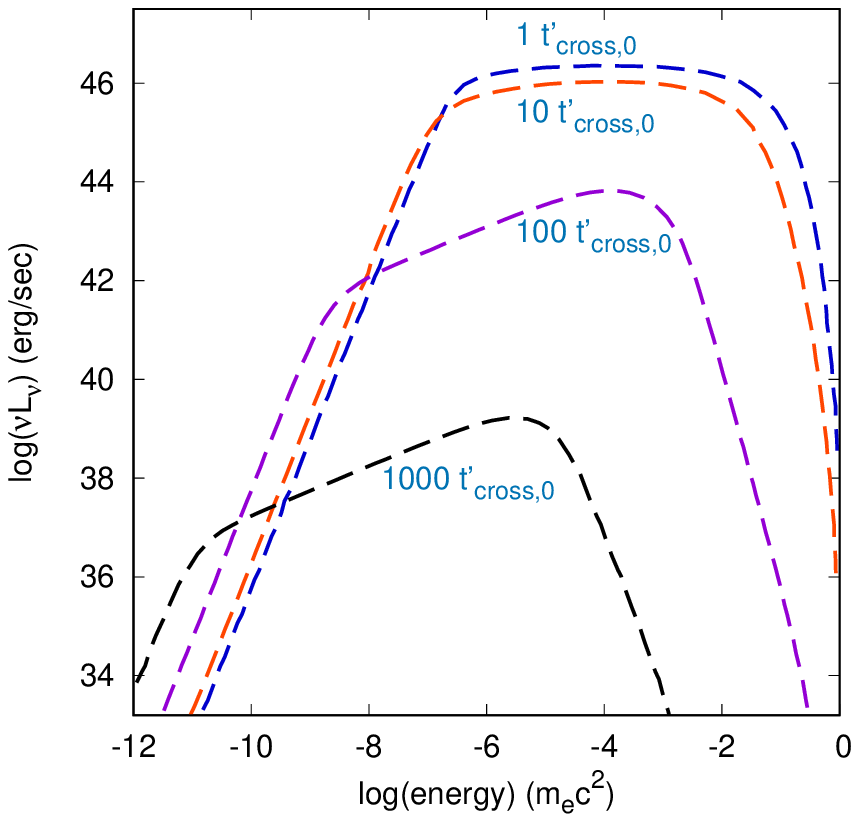}
\caption{}
\label{Fig:comp-photon}
\end{subfigure}

\caption{(a) Plot of the electron energy distribution at different time-steps for synchrotron and adiabatic losses only. The values of the parameters are  $R'_{\rm 0}=10^{15}$ cm, $L_{\rm e_{\rm 0}}^{\rm 'inj}=10^{42}$ erg s$^{-1}$, $u_{\rm exp}=0.1$ c, $B'_{\rm 0}=100$ G, $\gamma_{\rm min}=1$, $\gamma_{\rm max}=10^6$, $p=2$, $\delta=10$ and the profile of magnetic field strength decreases as $B'\propto R'^{-2}$ and the electron luminosity is constant in time. Full lines represent analytical solutions, while dashed lines the corresponding numerical ones. (b) Numerical results of the photon SED for the time instances of the depicted electrons.}%
\label{Fig:Comp}%
\end{figure*}
In order to calculate the electron evolution inside the blob, one needs to take into account, as
in the standard leptonic model, synchrotron and inverse Compton losses. However, in the present case, adiabatic losses should also explicitly be taken into account due to the expansion of the blob. One should also consider the possibility that the magnetic field inside the blob might change over time (or radius) for the same reason, affecting the synchrotron losses. Therefore we parametrize the magnetic field $B'$ as
\begin{equation}\label{eq:B}
B'(R')=B'_{\rm 0}\left({\frac{R'_{\rm 0}}{R'}}\right)^s,
\end{equation}
where $B'_{\rm 0}$ is the initial value of the magnetic field and $s$ is a free parameter. For the parameter $s$ we have $1\leq s \leq 2$ from magnetic flux
conservation with magnetic-field orientations between the extremes of purely toroidal and purely poloidal fields, respectively.

 Similarly, the electron injected power might also be a function of time and we parametrize it as 
\begin{equation}\label{eq:L}
L_{\rm e}^{\rm 'inj}(R')=\Leo\left( \frac{R'_{\rm 0}}{R'} \right)^{q},
\end{equation}
where $\Leo$ is the initial electron power and $q$  is a free parameter\footnote{The free parameters $s,~q$ are related to the jet dynamics, but in our investigation we do not restrict them to values from specific models.}.

We have developed a numerical code, based on \cite{MK95}, to calculate the evolution of the electron and photon distribution function in an expanding spherical source. This code solves two integro-differential equations, each describing the losses/sinks and injection of relativistic electrons and photons in the emitting region. The kinetic equation for the differential, with respect to energy, electron number $N_{\rm e}$ reads:
\begin{equation}\label{eq:ne}
\frac{\partial N_{\rm e}(\gamma,R')}{\partial t'}+\sum_{i} \frac{\partial}{\partial \gamma}\mathcal{L}_{\rm e}^{\rm i} N_{\rm e}(\gamma,R')+\frac{N_{\rm e}}{t'_{\rm esc,e}} =\sum_{i} Q_{\rm e}^{\rm i}(\gamma,~R'). 
\end{equation}
We denote by $\mathcal{L}_{\rm e}^{\rm i}$ the loss terms and by $Q_{\rm e}$ the injection terms of relativistic electrons. Also, $\gamma={(1-{u^2}/{c^2})}^{-1/2}$ the electron Lorentz factor with $u$ the electron velocity and $t'_{esc,e}$ the electron escape timescale. Specifically for the losses we write: 
\begin{equation}
\sum_{i}\frac{\partial}{\partial \gamma} \mathcal{L}_{\rm e}^{\rm i}   = \frac{\partial}{\partial \gamma}\left[\mathcal{L}_{\rm e}^{\rm syn}(\gamma,~R')+\mathcal{L}_{\rm e}^{\rm ICS}(\gamma,~R')+\mathcal{L}_{\rm e}^{\rm ad}(\gamma,~R')\right],
\end{equation}
where $\mathcal{L}_{\rm e}^{\rm syn}(\gamma,~R'), ~\mathcal{L}_{\rm e}^{\rm ICS}(\gamma,~R'), ~\mathcal{L}_{\rm e}^{\rm ad}(\gamma,~R')$ are the loss-rates for synchrotron emission, inverse Compton scattering\footnote{Here we present the losses in Thomson limit, however the code uses also a numerical treatment for Klein-Nishina losses as given by \cite{B&G}.}, and adiabatic expansion which are given by the expressions:
\begin{equation}\label{eq:los:syn}
  \mathcal{L}_{\rm e}^{\rm syn}(\gamma,R')=- \dfrac{4}{3}\sigma_{\tau}c\frac{U'_{\rm B}(R')}{m_{\rm e} c^2} \beta^2 \gamma^2,  
\end{equation}
\begin{equation}\label{eq:los:ics}
 \mathcal{L}_{\rm e}^{\rm ICS}(\gamma,R')=-\dfrac{4}{3}\sigma_{\tau}c\frac{U'_{\rm phot}(R')}{m_{\rm e} c^2} \beta^2 \gamma^2,   
\end{equation}
 and
 \begin{equation}\label{eq:los:ad}
    \mathcal{L}_{\rm e}^{\rm ad}(\gamma,R')= -\dfrac{u_{\rm exp}}{R'}\gamma,
 \end{equation} 
 with $\beta=u/c$, $U'_{\rm B}(R') $ and $U'_{\rm phot}(R') $ the energy densities in magnetic fields and photon fields, respectively. 
 \begin{figure*}
\centering
\includegraphics[width=17cm]{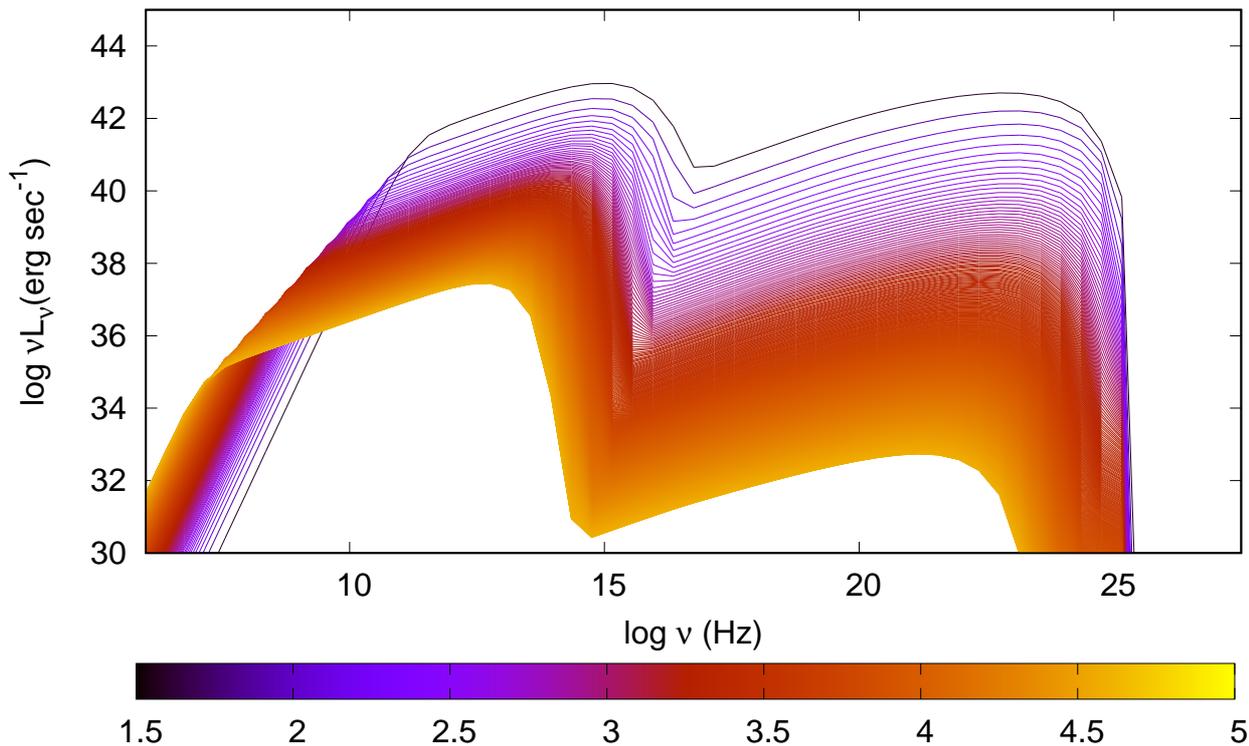}
\caption{The time evolution of the SED when all the physical processes are taken into account. The parameters have the values: $R'_{\rm 0}=10^{15}$ cm, $B_{\rm 0}=3.16$ G, $L_{\rm e_{\rm 0}}^{\rm 'inj}=10^{43}$ erg s$^{-1}$, $t'_{\rm esc,e}=R'/c$, $u_{\rm exp}=0.1$ c, $\gamma_{\rm min}=1$, $\gamma_{\rm max}=10^{4.5}$, $p=2$ and $\delta=10$.  The magnetic field and electron luminosity profiles are $\propto R'^{-1}$. The colorbar refers to time, the time unit is given in days in the co-moving frame} and it is in a logarithmic scale. 

\label{Fig:timeSED}
\end{figure*}
The injection of relativistic electrons is given by 
\begin{equation}
 \sum_{i} Q_{\rm e}^{\rm i}(\gamma,~R')= Q_{\rm e}^{\rm inj}(\gamma,~R') +Q_{\rm e}^{\gamma\gamma}(\gamma,~R'), 
\end{equation}
where 
 $Q_{\rm e}^{\rm inj}(\gamma,~R')$ is the injection term of primary accelerated electrons and 
$Q_{\rm e}^{\gamma \gamma}(\gamma,~R')$ is the injection of the pairs that are created by the photon-photon absoprtion. 
For the injected electron term $Q_{\rm e}^{\rm inj}$ we assume that it is related to the electrons injected power by the relation:
\begin{equation}
L_{\rm e}^{\rm 'inj}(R')=(m_{\rm e}c^2)^2\int_{\gamma_{\rm min}}^{\gamma_{ max}} Q_{\rm e}^{\rm inj}(\gamma, R') \gamma {\rm{d}}\gamma,
\end{equation}
where  we have assumed a power-law injection
\begin{equation}
Q_{\rm e}^{\rm inj}(\gamma,R')= q_{\rm e}(R') \gamma^{-p} = q_{\rm e,0} \left(\frac{R'_{\rm 0}}{R'}\right)^{q}\gamma^{-p},\quad \gamma_{\rm min}\leq \gamma\leq\gamma_{\rm max},
\end{equation}
with  $\gamma_{\rm min}$, $\gamma_{\rm max}$ the minimum and maximum Lorentz factors of the electron distribution, respectively.\\
The kinetic equation for the differential, with respect to energy, photon number $N_{\rm \gamma}$ has the form:
\begin{equation}\label{eq:ng}
\frac{\partial N_{\rm \gamma}(x',R')}{\partial t'}+ \frac{N_{\rm \gamma}(x',R')}{t_{{\rm cross}}'}+\sum_{i} \mathcal{L}_{\rm \gamma}^{\rm i}N_{\rm \gamma}(x',R') =\sum_{i} Q_{\rm \gamma}^{\rm i}(x',~R'), 
\end{equation}
where $x$ is the dimensionless photon frequency, {\sl i.e.} $x'=h\nu'/(m_{\rm e} c^2)$. Photons escape from the source on the  timescale $t'_{{\rm cross}}=\dfrac{R'}{c}$ which is the light crossing time of the source at time $t'$ and it is measured in units of the initial light crossing time 
\begin{equation}
t'_{\rm cross,0}=\frac{R'_{\rm 0}}{c}.
\end{equation}

The processes of photon-photon absorption and synchrotron self-absorption contribute to the photon sinks, 
\begin{equation}
    \sum_{i} \mathcal{L}_{\rm \gamma}^{\rm i}(x',~R')~=~\mathcal{L}_{\rm \gamma}^{\gamma\gamma}(x',~R')+\mathcal{L}_{\rm \gamma}^{\rm ssa}(x',~R'),
\end{equation}
where $\mathcal{L}_{\rm \gamma}^{\gamma\gamma}(x',~R')$ and $\mathcal{L}_{\rm \gamma}^{\rm ssa}(x',~R')$
denote the processes of photon-photon absorption and synchrotron self-absorption\footnote{In our code we use the full expression in the calculation of synchrotron self-absorption in contrast to the $\delta$-function approach that was used in \cite{MK95}.} respectively. The photon source function is given by
\begin{equation}
   \sum_{i} Q_{\rm \gamma}^{\rm i}(x',~R')~=~Q_{\rm \gamma}^{\rm syn}(x',~R')+Q_{\rm \gamma}^{\rm ICS}(x',~R') 
\end{equation}
 where $Q_{\rm \gamma}^{\rm syn}(x',~R')$ and $Q_{\rm \gamma}^{\rm ICS}(x',~R')$  are the terms corresponding to synchrotron\footnote{Here we use the full expression in our calculations in contrast to the $\delta$-function approach that was used in \cite{MK95}.} radiation and to inverse Compton scattering, respectively.
 
 Equation \ref{eq:ng} by construction calculates the differential number of photons inside the source when its radius is $R'$. From here, one can calculate the co-moving spectral luminosity (erg/sec/Hz) of the source using 
 \begin{equation}
    L_{\rm \nu'}=\frac{3N_{\gamma}(x',R')}{t'_{\rm cross}(R')}\left( \frac{{x' m_{\rm e}c^2}}{h}\right)
 \end{equation}
 and then calculate the observed flux by imposing the usual Doppler transformations. Therefore, Eq. \ref{eq:ng} calculates an instantaneous flux coming from a blob of radius $R'$. 

In order to check the numerical scheme presented above, especially as far as the expansion is concerned -- the various rates have already been tested, see \cite{MK95} and more recently in \cite{Cerutti2021}, we solve the kinetic equation for the electrons (Eq. \ref{eq:ne}) analytically by taking into account only synchrotron and adiabatic losses and we compare the solution to the results of the code.  Figure \ref{fig:comp-elec} depicts the electron distribution function in the case of continuous electron injection for a decreasing magnetic field. The electron number constantly increases with time as they continuously accumulate inside the source, while the shape of the distribution changes continuously. At early times the distribution is completely cooled due to strong synchrotron losses, however, at later times, adiabatic losses take over and the cooling break moves progressively to higher energies. The analytical and numerical solutions are identical, with the only difference appearing at high energies since we have not included cutoff effects in our analytical solution.
Figure \ref{Fig:comp-photon} depicts the corresponding photon spectrum, which is calculated numerically. The luminosity decreases with time since cooling becomes less efficient at later times while the spectral turnovers correspond to the electron cooling breaks shown in Fig. \ref{fig:comp-elec}. 
 
Figure \ref{Fig:timeSED} shows an example when all processes are taken into account in the numerical code. It depicts the time evolution of the SED when the source constantly expands, while the magnetic field strength and the electron injection luminosity decrease with time, i.e., $q=s=1$.  The addition of all processes complicates the evolution over the case of Fig. \ref{Fig:Comp}, however, some basic features remain the same. The photon luminosity decreases with time, represented by the color gradient in the logarithmic scale. The SED has two characteristics bumps; the low-frequency component is produced by synchrotron radiation while the higher one is produced due to Synchrotron Self-Compton (SSC). The synchrotron frequency peak is given by the expression $\nu_{\rm syn}^{\rm 'pk}\propto B'(R') \gamma_{\rm max}^2$, so as the magnetic field strength decreases, the peak shifts to lower energies. The same behavior is shown by the Compton peak, which depends both on the energy of synchrotron photons and on $\gamma_{\rm max}$ through SSC mechanism. Also, the SSC luminosity decreases quadratically compared to the synchrotron one. Moreover, the spectrum shows a break at radio frequencies due to synchrotron self-absorption, which shifts to lower energies as time increases and the source expands. The slope of the spectrum below the synchrotron self-absorption frequency has the characteristic value of $\nu F_{\nu}\propto \nu^{7/2}$. At all frequencies, the spectral flux peaks at the beginning of the expansion, apart from the radio band which shows a peak at later times. The produced luminosity 
at these frequencies
increases with time, reaches a peak at the transition from the optically thin to the optically thick regime and then decreases.  We will expand on this issue in the next session.

\section{Localization of the onset of radio emission}\label{s3}
\indent

\begin{figure}[!htbp]
\centering
\includegraphics[width=8cm]{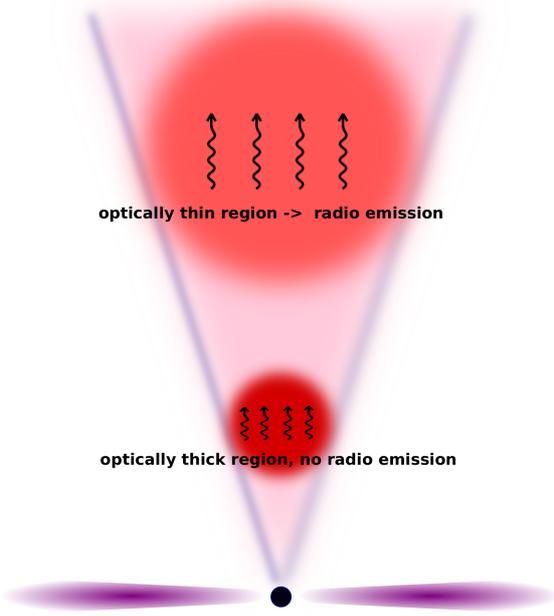}
\caption{A schematic representation of the one-zone expanding model. The emitting region is optically thick to radio emission close to the central engine. As it propagates along the axis of the jet and expands, it transits to the optically thin region. On the other hand, higher frequencies are emitted from all distances.}
\label{Fig:Model}
\end{figure}

Very-long-baseline interferometry (VLBI) studies have given much information about the structure of the jets and specifically about pc-scale jets.
As mentioned in the previous section, we can detect radio emission mainly from the regions where the source is optically thin at radio 
frequencies. In this section, we investigate the role of the expansion velocity on the transition from the optically thick to the optically thin region in the case of an expanding blob, see Fig. \ref{Fig:Model}. 
%-----------------------------------------------------------------
% \indent
In this case, the frequency below which the synchrotron radiation is absorbed can be derived by the condition for the optical depth $\tau_{\rm \nu'_{ssa}}=\alpha'_{\rm \nu'_{ssa}}R' \sim 1$ where $\alpha'_{\rm \nu'_{ssa}}(t')$ is the absorption coefficient, e.g., Eq. 6.53 from \cite{RL1979}.
The synchrotron self-absorption frequency $\nu'_{\rm ssa}$ depends on the specifics of the electron distribution and on the magnetic field. In our treatment, both of these depend on the radius of the blob and therefore on the co-moving time, see Eqs. \ref{eq:B} and \ref{eq:L}.
 \begin{figure}[!htbp]
%	\addtocounter{figure}{-1}
\centering
\includegraphics[width=10cm]{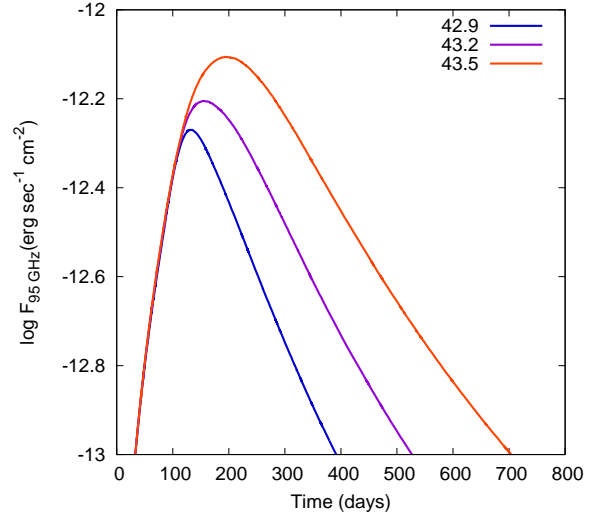}
\caption{Radio light-curves as a function of time as measured in the observer's frame for various initial values in $\log \Leo $ (erg/sec) of the injected electron
luminosity -- the injection is assumed to start at $t=0$.
The other parameters have the values: $R'_{\rm 0}=10^{16.5}$ cm, $u_{\rm exp}=0.05$ c, $B'_{\rm 0}=1$ G, $t'_{\rm esc,e}=R'/c$, $\gamma_{\rm min}=1$, $\gamma_{\rm max}=10^6$, $p=2$ and $\delta=10$. For the magnetic field profile, we have $B'\propto R'^{-1}$ and for the electron luminosity  $L_{\rm e}^{\rm 'inj}\propto R'^{-2}$. The source is assumed to be at a distance of 130 Mpc.}
\label{fig:init_95}
\end{figure}
Therefore, following the assumptions above, the expression of synchrotron self-absorption for our numerical model has the form\footnote{Here, we neglect energy losses and electron escape from the source.}: 

\begin{equation*}
    \nssa(R') = \left[k_{\rm 1} k_{\rm 2}(p)\frac{\Leo B_{\rm 0}^{'\frac{p+2}{2}}}{u_{\rm exp}}\frac{1}{R^{'2}}\left(\frac{R'_{\rm 0}}{R'}\right)^{\frac{s(p+2)}{2}}\right]^{\frac{2}{p+4}}
\end{equation*}
\begin{equation}\label{Eq:nssaRL}
~~~~~~~~~~~~~~~~~~~~~~~~~~~~~~\times
\left\{ \begin{array}{cc}
                 \left[R'_0\ln\left(\frac{R'}{R'_0}\right)\right]^{\frac{2}{p+4}} &  {\rm for} \ q=1 \\  
                 \phantom{} & \phantom{} \\
                 \left[\frac{R'_0}{1-q}\left[\left(\frac{R'_0}{R'}\right)^{q-1}-1\right]\right]^{\frac{2}{p+4}} & {\rm for} \ q \neq 1
                 
                \end{array}
\right.
\end{equation}

% \begin{equation}\label{Eq:nssaRL}
% \nssa(R')= \left[k_{\rm 1} k_{\rm 2}(p)\frac{\Leo B_{\rm 0}^{'\frac{p+2}{2}}}{u_{\rm exp}}\frac{\int_{R'_{\rm 0}}^{R'} \left(\frac{R'_{\rm 0}}{R'}\right)^q \rm{d}R'}{R^{'2}}\left(\frac{R'_{\rm 0}}{R'}\right)^{\frac{s(p+2)}{2}}\right]^{\frac{2}{p+4}}
% \end{equation}
where 
\begin{equation*}
k_{\rm 1}=\frac{3\sqrt{3}q^3}{32\pi^2 m_{\rm e}} 
\end{equation*}
and 
\begin{equation*}
k_{\rm 2}(p)=\left(\frac{3q}{2\pi m^3 c^5}\right)^{\frac{p}{2}}\Gamma\left(\frac{3p+2}{12}\right)\Gamma\left(\frac{3p+22}{12}\right)\sin^{\frac{p+2}{2}}\alpha,
\end{equation*}
where $\Gamma(y)$ is the Gamma function.
The luminosity at the synchrotron self-absorption frequency becomes, e.g., \cite{Stawarz08}:  
% \begin{equation*}
% [\nu' L'_{\nu}]_{\rm syn,ssa}=\frac{c\sigma_{\rm T}}{12\pi m_{\rm e}c^2}\frac{B^{'2}(R')}{u_{\rm exp}} \frac{\int_{R'_{\rm 0}}^{R'} L_{\rm e}^{\rm 'inj}(R')\rm{d}R'}{\int_{\gamma_{\rm min}}^{\gamma_{\rm max}}\gamma^{1-p}\rm{d}\gamma}\left(\frac{4\pi m_{\rm e} c \nssa(R')}{3 eB'(R')}\right)^{\frac{3-p}{2}},
% \end{equation*}
\begin{equation}\label{eq:s08}
[\nu' L'_{\nu}]_{\rm syn,ssa}=\frac{c\sigma_{\rm T}}{12\pi m_{\rm e}c^2}\frac{B^{'2}(R')}{u_{\rm exp}} \frac{\int_{R'_{\rm 0}}^{R'} L_{\rm e}^{\rm 'inj}(R)\rm{d}R}{\int_{\gamma_{\rm min}}^{\gamma_{\rm max}}\gamma^{1-p}\rm{d}\gamma}\left(\frac{4\pi m_{\rm e} c \nssa(R')}{3 eB'(R')}\right)^{\frac{3-p}{2}},
\end{equation}
where $\sigma_{\rm T}$ is the Thomson cross-section.

 As Eqs. \ref{Eq:nssaRL} and \ref{eq:s08} show, both the synchrotron self-absorption frequency and its corresponding luminosity depend on the initial values and profiles of the key parameters. As we show below, for all physical combinations of these, the source starts self-absorbed and at some point, it becomes transparent. Practically speaking, the distance $z_{\rm ssa}(\nu')$ where this occurs marks the onset of radio emission at frequency $\nu'$ and corresponds to the maximum of the lightcurve at that frequency, which can be found from Eq.~\ref{eq.zt} at a time $t'_{\rm ssa}(\nu')$ defined by $\tau_{\rm ssa}(t'_{\rm ssa}(\nu'))=1$. Therefore for
$z>z_{\rm ssa}(\nu')$ the source is optically thin.
\begin{figure}[!htbp]
%	\addtocounter{figure}{-1}
\centering

\begin{subfigure}[b]{0.45\textwidth}
\centering
\includegraphics[width=9cm,trim=60 4 4 6]{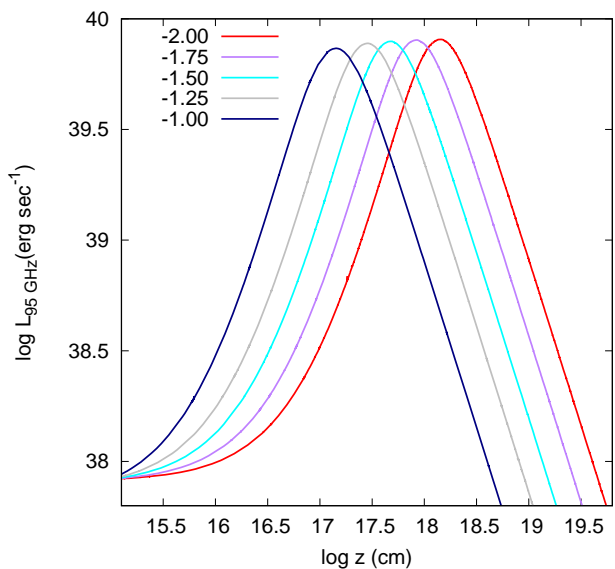}
\caption{}
\label{fig:sub-locu}
\end{subfigure}
\hspace{1cm}
\begin{subfigure}[b]{0.45\textwidth}
\centering
\includegraphics[width=9cm, trim=60 4 4 6]{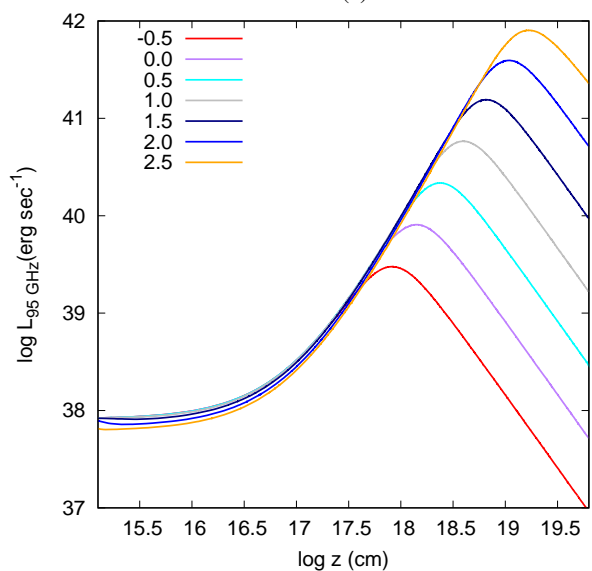}
\caption{}
\label{fig:sub-locB}
\end{subfigure}

\caption{The localization of the onset of the radio emission depending on (a) the expansion velocity in units of c and (b) the initial magnetic field strength, both in logarithmic scale. The other parameters have the values: $R'_{\rm 0}=10^{15}$ cm, $L_{\rm e_{\rm 0}}^{\rm 'inj}=10^{42}$ erg s$^{-1}$, $t'_{\rm esc,e}=R'/c$, $\gamma_{\rm min}=1$, $\gamma_{\rm max}=10^6$, $p=2$, $\delta=10$ and $z_{\rm 0}=0.001$ pc. In sub-figure (a) $B'_{\rm 0}=1$ G and in sub-figure (b) $u_{\rm exp}=0.01$ c. The profiles of the magnetic field strength and electrons luminosity are decreasing as $R^{-1}$}.%
\label{Fig:Localization}%
\end{figure}

An example of the above can be seen in 
Fig. \ref{fig:init_95} that depicts the radio lightcurve at 95 GHz\footnote{When we use the symbols $F_i,~L_i$ where $i$ a specific photon energy/frequency, we present the flux/luminosity for this specific energy/frequency, respectively.} as a function of time (as measured on Earth) for three  different values of the injected electron luminosity in $\log \Leo $ (erg/sec).
 As we mentioned in Fig. \ref{Fig:timeSED}, the radio luminosity increases initially with time, it reaches a peak during the transition from the optically thick to the optically thin regime and then decreases. The transition occurs later and is brighter for higher electron luminosities because the number of electrons increases with luminosity. This result is in qualitative agreement with Eqs. \ref{Eq:nssaRL} and \ref{eq:s08}. As it turns out,
 the slopes of the lightcurves depend on the profile of the magnetic field strength (e.g., for $B'\propto R^{'-2}$ the decay is faster in comparison with $B'\propto R^{'-1}$)  and of the electron luminosity. 

Figure \ref{Fig:Localization} depicts the luminosity at 95 GHz as a function of z when all parameters are kept constant except for the expansion velocity (Fig.~\ref{fig:sub-locu}) and the initial value of the magnetic field (Fig.~\ref{fig:sub-locB}). Figure~\ref{fig:sub-locu} shows that as the velocity increases, the transition from the optically thick to the optically thin regime occurs at smaller distances, a result that is expected from Eq. \ref{Eq:nssaRL}. Qualitatively this is to be expected as all the physical quantities change faster; therefore the transition occurs closer to the
central engine. 

Figure \ref{fig:sub-locB} that examines the effects of the blob's initial magnetic field $B'_{\rm 0}$ on the transition distance $z_{\rm ssa}$ shows that the transition to the optically thin regime occurs at higher distances as $B'_{\rm 0}$ increases, while the luminosity during the transition also increases. It is interesting to note that transition occurs around the parsec scale for the most commonly assumed parameters. Finally, we should also mention that we find a very similar behavior if we were to increase, instead of $B'_{\rm 0}$, the initial electron luminosity. 

\begin{figure*}[!htbp]
%	\addtocounter{figure}{-1}
\centering

\begin{subfigure}[b]{0.45\textwidth}
\centering
\includegraphics[width=8.5cm,trim=60 4 4 6]{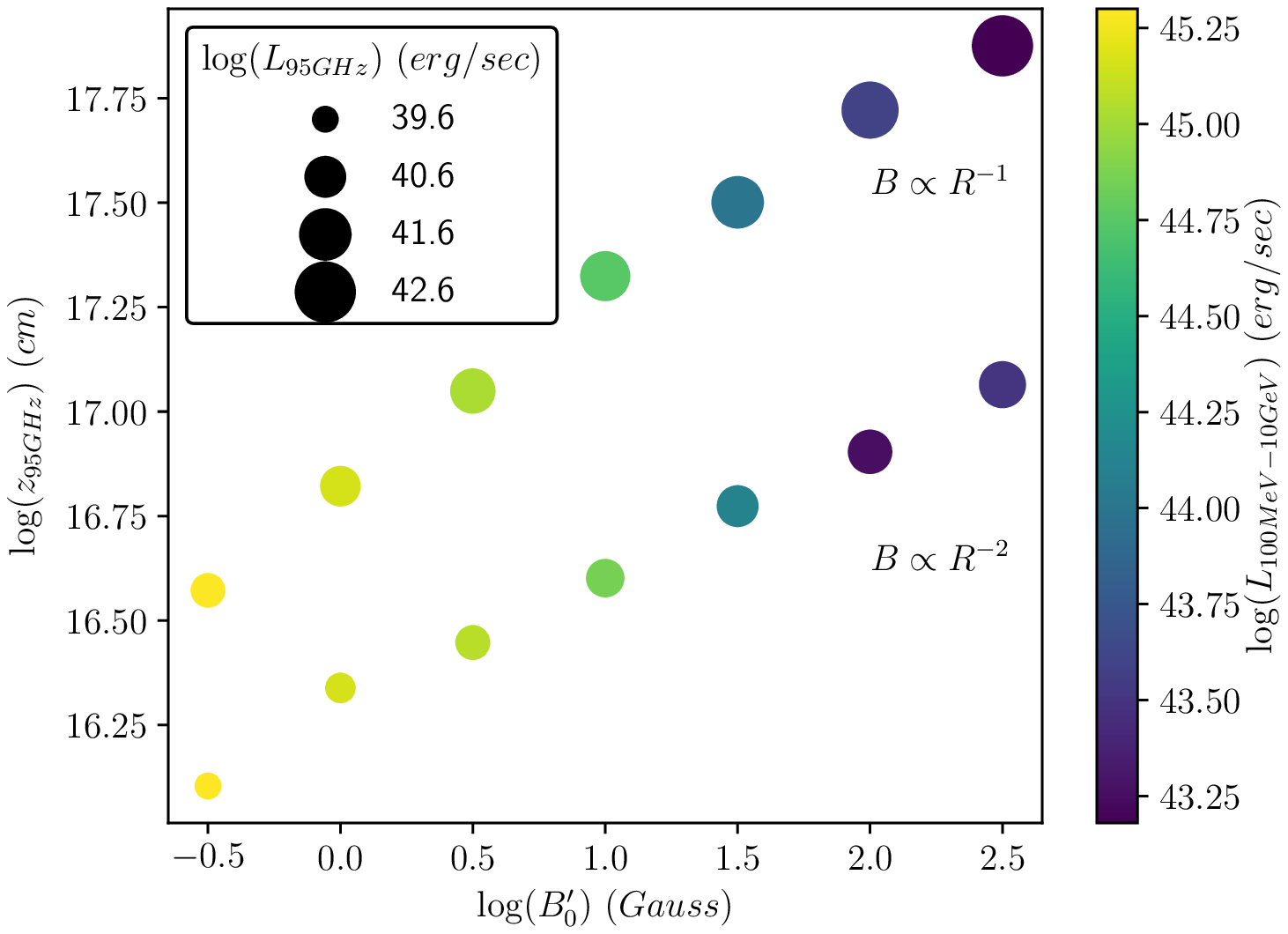}
\caption{}
\label{fig:sub-BL}
\end{subfigure}
\hspace{1cm}
\begin{subfigure}[b]{0.45\textwidth}
\centering
\includegraphics[width=8.5cm, trim=60 4 4 6]{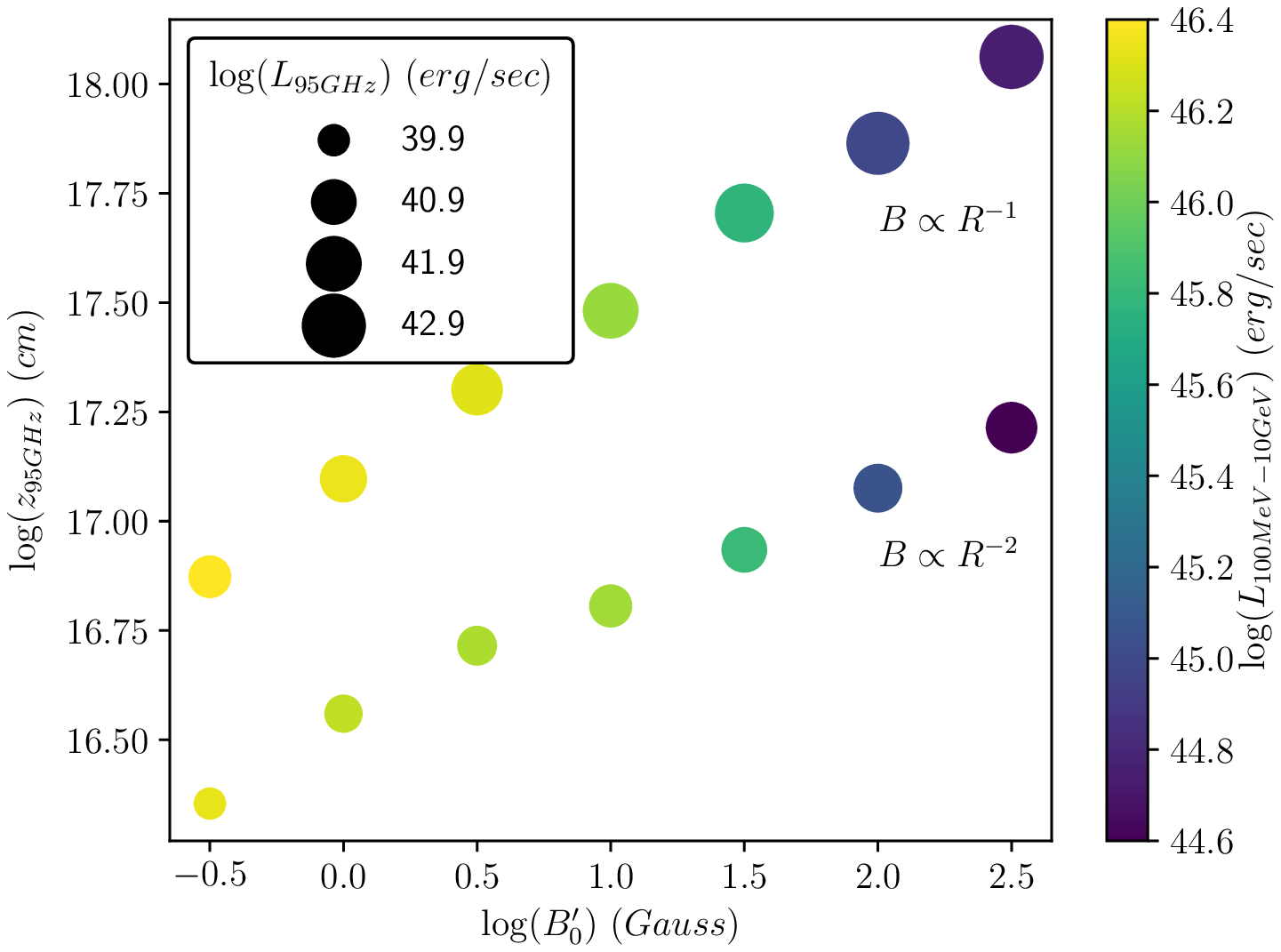}
\caption{}
\label{fig:sub-B10L}
\end{subfigure}

\caption{The dependence on initial magnetic field value of the distance where the radio photons escape. The figures depict the two different profiles of the magnetic field strength depending on radius. Also, the peak of $\gamma$-ray -- at the base of the jet-- and the peak of radio luminosity -- at the distance where the source becomes optically thin, $z_{ssa}(95 GHz)=z_{95~{\rm GHz}}$ -- are represented. The electron power have two different values: (a) $L_{\rm e_{\rm 0}}^{\rm 'inj}=10^{42}$ ${\rm erg ~s^{-1}}$, (b) $L_{\rm e_{\rm 0}}^{\rm 'inj}=10^{43}$ ${\rm erg ~s^{-1}}$. The other parameters have the values: $R'_{\rm 0}=10^{15}$ cm, $t'_{\rm esc,e}=R'/c$, $u_{\rm exp}=0.1$ c, $\gamma_{\rm min}=1$, $\gamma_{\rm max}=10^6$, $p=2$, $\delta=10$ and $z_{\rm 0}=0.001$ pc. The profile of the electron power follows the magnetic field one.}%
\label{fig:Mag}%
\end{figure*}

\section{Simulating flaring episodes in blazars}\label{s4}
Blazars are characterized by their variability across the electromagnetic spectrum on time scales from minutes to years. Every flaring episode has its characteristic signature, a fact that makes its modeling a challenge. There are many studies in the literature (e.g. \cite{1992Camenzind,1995Wagner,2013Bhatta,M14}) based on the flare's dependence on the accretion disk and jet plasma variations. Here, we use a simplified way to simulate flaring activity and we present two approaches: radio flares produced due to the expansion of the blob and flares, at all frequencies, produced due to particle reaccelerating episodes. In this section we will not focus only radio flares, but we will examine also higher frequencies. 

\subsection{Flares by energetic blobs}
The continuous injection of a series of 
expanding blobs could create emissions at all z's. If these blobs had similar initial conditions, then the superposition of their emission at all distances could create a steady-state, quiescent jet (e.g., \cite{Potter18,Boula19b}). Suppose now that one of the blobs has different initial properties than the rest that make the steady-state emission. This will create a disturbance in the emission that will propagate with the particular blob down the jet.
In the case where the initial parameters of this disturbance are such that it could lead to higher emission than that of the surrounding blobs, then a flaring episode will ensue and one could calculate the time-dependent emission from it.

According to our analysis in Section \ref{s2}, for conditions where both the electron injection luminosity and the magnetic field drop outwards (i.e. $q>0$ and $s>0$), the high-energy photons show a peak very early, close to the base of the jet  -- see Fig. \ref{Fig:timeSED}.
This occurs because the choice of the parameters makes radiation losses more important at the beginning of the expansion. At radio frequencies, however,
the peak of the corresponding lightcurve occurs at much larger distances when the blob has reached the region of the transition from the optically thick to the optically thin regime -- see Section \ref{s3}. The details of the shape of the radio flare will depend on many parameters, such as the magnetic field strength and the electrons' power profile on the distance.
These parameters, along with the expansion velocity, determine in principle the time-lags between higher frequencies which show a maximum essentially at the base of the jet and the radio frequencies which have their maximum much further out \citep[see also][]{Boula19b}. 

The above can be encoded in Fig. \ref{fig:Mag} which depicts the distance where the peak of the radio flare occurs as a function of the initial magnetic field value for two magnetic field profiles and two initial electron injections luminosities. The size of the circles scales with the peak radio luminosity at 95 GHz, while the color encodings shows the peak of $\gamma$-ray luminosity between 100 MeV and 10 GeV\footnote{In the present analysis, high-energy photons are produced by synchrotron self-Compton; thus, these results refer mainly to BL Lac objects, where the external photon fields are negligible.} -- note that the $\gamma$-ray peak is not contemporaneous to the peak of the radio flare as it 
has occurred at earlier times, soon after the injection of the energetic blob. 

Several conclusions can be drawn from  Fig. \ref{fig:Mag}. Radio flares tend to be more energetic and occur at larger distances along the jet when $B'_{\rm 0}$ is higher -- results already drawn from Fig.~\ref{fig:sub-locB}. Furthermore, an anti-correlation between the peak luminosities in $\gamma$-ray and radio is found. This is to be expected because, for the same injected electron luminosity, higher values of $B'_{\rm 0}$ tend to suppress the $\gamma$-ray which are produced by the SSC process. Furthermore, the source becomes optically thin at smaller distances when the magnetic field strength and electron power decrease faster with radius. 
Finally, when the initial power of the injected electrons becomes higher, then all quantities increase but in a slower than linear way -- compare Figs.~\ref{fig:sub-BL} and \ref{fig:sub-B10L}.

\subsection{Flares by reaccelerating episodes}

An alternative way for producing a flare is to consider a reaccelerating episode in an expanding blob when this is at some distance $z_*$ from the origin. As in the previous sections, we have not included an acceleration mechanism; we treat, instead, this case by increasing the injected power of the relativistic electrons through a Lorentzian function in time:
\begin{equation*}
Q_{\rm e}(\gamma,t')= q_{\rm e}(t') \gamma^{-p} \left(1+\frac{\alpha w^{'2}}{4(t'-t'_*)^2+w^{'2}}\right)
\end{equation*}
\begin{equation}\label{eq:loretzian}
= q_{\rm {e,0}} \left( \frac{R'_{\rm 0}}{R'} \right)^{q}\gamma^{-p}\left(1+\frac{\alpha w^{'2}}{4(t'-t'_*)^2+w^{'2}}\right), \quad \gamma_{\rm min}\leq \gamma\leq\gamma_{\rm max}
\end{equation}
where $\alpha$ is the value at maximum, $w'$ the width of the injection and $t'_*$ the time of the maximum which corresponds to $z_*$. The lorentzian form of the injection is partly motivated by the shape of the pulses which are observed \citep[see also][]{2001Sikora,2002Wang}.

This assumption allows us to examine the impact of our free parameters especially, the expansion velocity on the shape of the flares and to study the interplay between radiative and adiabatic losses. 
\begin{figure}[!htbp]
\centering
\includegraphics[width=9cm, trim=60 4 4 6]{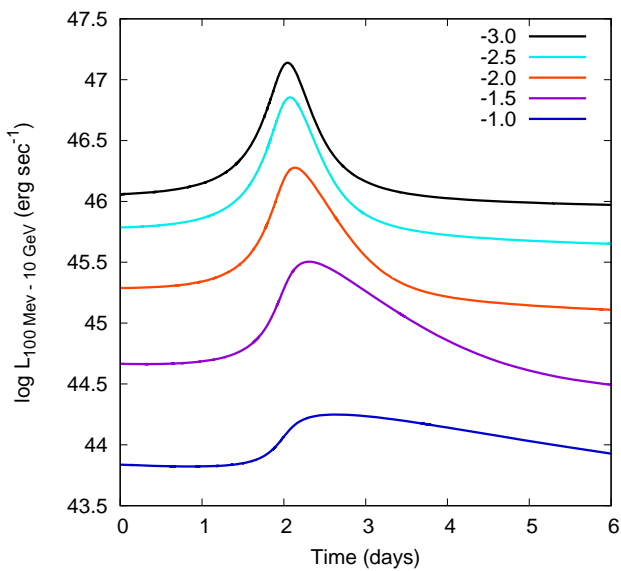}
\caption{Flares in $\gamma$-ray produced by reaccelerating episodes in blobs which have the same initial parameters but different expansion velocities -- these are given in units of c in a logarithmic scale. All flaring parameters (see Eq. \ref{eq:loretzian}) are the same.
As the velocity increases, the symmetry of the flare breaks as the decay time becomes progressively longer than the rise time producing an extended flare.
}
\label{Fig:u}
\end{figure}
\begin{figure}[!htbp]
\centering
\includegraphics[width=9cm, trim=60 4 4 6]{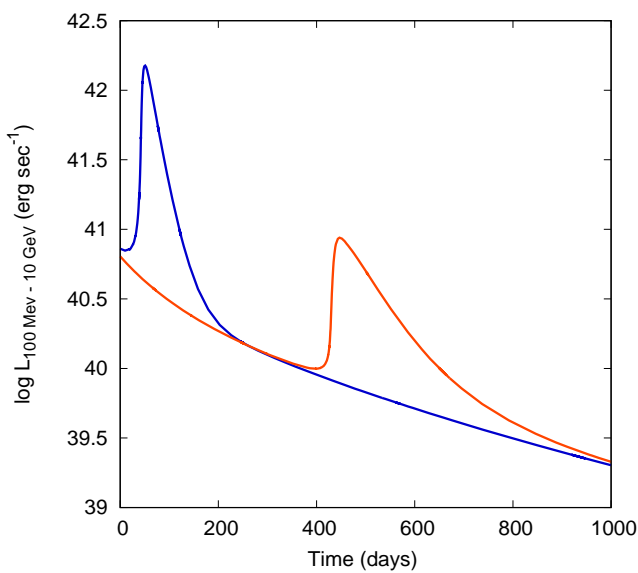}
\caption{Flares in $\gamma$-ray produced by a late reaccelerating episode occurring at distances $z_{\rm 1}=1$ pc (blue line) and  $z_{\rm 2}=10$ pc (red line).
The other parameters have the values:  $B'_{\rm 0}=10$~G, $R'_{\rm 0}=10^{16}$~cm, $L_{\rm e_{\rm 0}}^{\rm 'inj}=10^{44}$ erg s$^{-1}$, $t'_{\rm esc,e}=R'/c$, $u_{\rm exp}=0.1$ c, $\gamma_{\rm min}=1$, $\gamma_{\rm max}=10^5$, $p=2$, $\delta=10$ and the profiles of the magnetic field strength and electrons luminosity are decreasing as $R^{-1}$. The flaring parameters (see Eq. \ref{eq:loretzian}) are $w'=100$ $t_{\rm cross,0}$, and $\alpha=$100.
The flare 
which occurs earlier (blue line) is dominated by strong radiation losses, and it appears more symmetrical, contrary to the later one (red line), which is dominated by the adiabatic losses and electron escape. 
}
\label{Fig:Fl}
\end{figure}
Figure \ref{Fig:u} shows the role of $u_{\rm exp}$ on the pulse shape. Here all parameters remain constant between cases except of $u_{\rm exp}$. 
The lightcurves depict the behavior of $\gamma$-ray in a reaccelerating episode which occurs at a certain $t'_*$ which is the same for all cases. This means that episodes with smaller velocities produce flares closer to the origin, where both the magnetic field strength and the electron power are larger. This explains the fact that the $\gamma$-ray luminosity is decreasing with increasing $u_{\rm exp}$. At the same time, 
small expansion velocities produce more symmetrical flares because the cooling timescale $t'_{\rm cool} \ll t'_{\rm dyn}=t'_{\rm esc,e}=\frac{R'}{c}$, where $t'_{\rm dyn}$ is the dynamical timescale. On the other hand, fast expansion has a large impact on the cooling timescale because the magnetic field has dropped substantially as the reaccelerating episode occurs at larger distances and, as a result, the flare appears both asymmetrical and extended. 
\begin{figure*}[!htbp]
%	\addtocounter{figure}{-1}
\centering
\begin{subfigure}[b]{0.45\textwidth}
\centering
\includegraphics[width=9cm,trim=60 4 4 6]{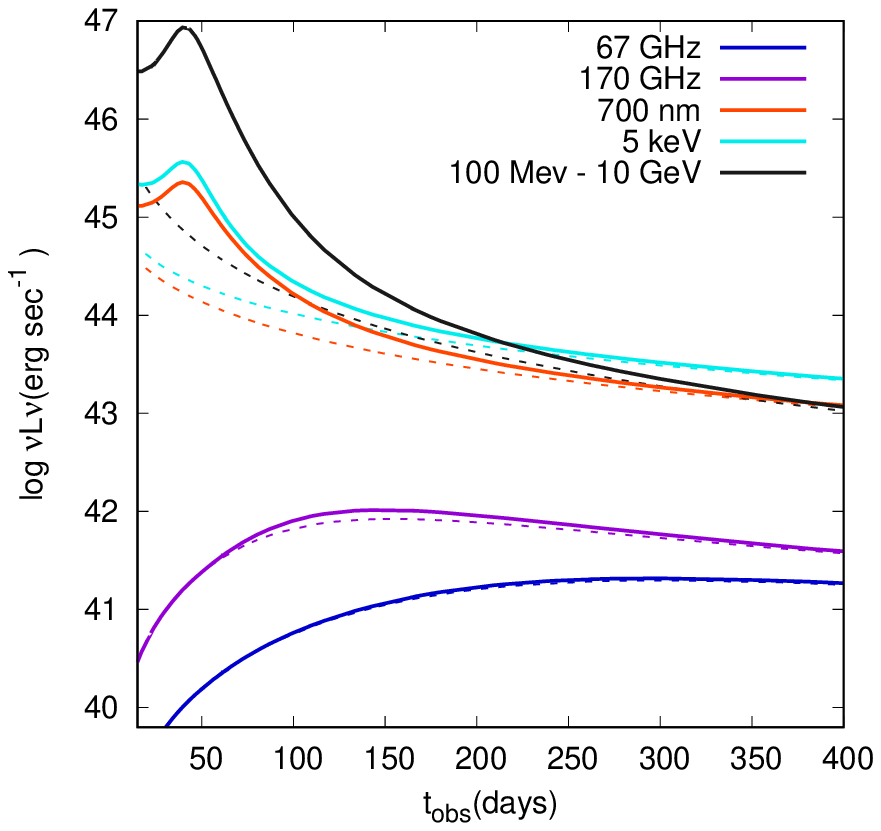}
\caption{}
 \label{fig:thicknessa}
\end{subfigure}
\hspace{1cm}
\begin{subfigure}[b]{0.45\textwidth}
\centering
\includegraphics[width=9cm, trim=60 4 4 6]{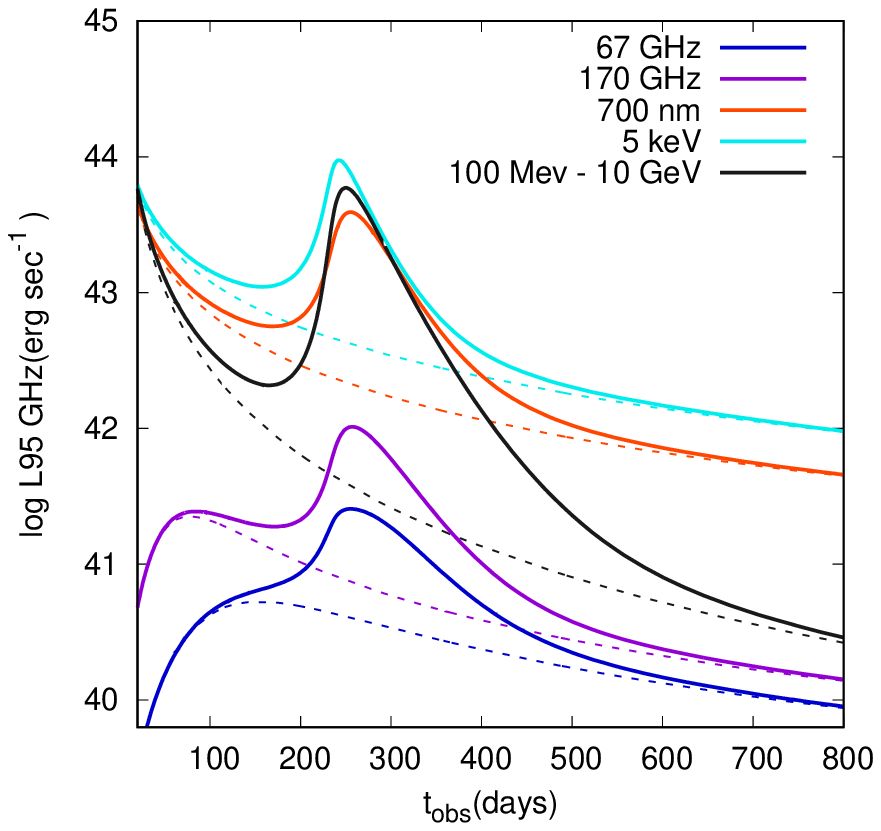}
 \caption{}
 \label{fig:thicknessb}
\end{subfigure}
\caption{(a) A flare in a self-absorbed region. The parameters have the values: $R'_{\rm 0}=10^{15}$ cm, $B'_{\rm 0}=10$ G, $L_{\rm e_{\rm 0}}^{\rm 'inj}=10^{43}$ erg s$^{-1}$, $t'_{\rm esc,e}=R'/c$, $u_{\rm exp}=0.1$ c, $\gamma_{\rm min}=1$, $\gamma_{\rm max}=10^{6}$, $p=2$, $\delta=10$ and $z_{\rm 0}=0.001$ pc. The magnetic field decreases as $B'\propto R'^{-1}$ and the electron luminosity as $L_{\rm e_{\rm 0}}^{\rm 'inj}\propto R'^{-1}$. For the $\gamma$-ray flare, we used a pulse which was injected at $t'_{*}=100~t_{\rm cross,0}$, with a width $w'=50~t_{\rm cross,0}$ and $\alpha=45$. The dashed lines represent the emission of the blob without the reaccelerating episode. (b) A  flare in an optically thin region. Here we change only the initial electron luminosity  $L_{\rm e_{\rm 0}}^{\rm 'inj}=10^{42}$  erg s$^{-1}$ and the reaccelerating episode occurs at the time $t'_{*}=600~t_{\rm cross,0}$ --the rest parameters are the same with case (a).}%
\label{fig:thickness}%
\end{figure*}
The same conclusion can be drawn from 
Fig. \ref{Fig:Fl}.  Here initial conditions and flaring parameters are identical except for the location of the peak of the reaccelerating episode -- see Eq. \ref{eq:loretzian}. The flare that occurs closer to the origin (blue line) is more symmetric than the one that occurs further away (red line), which is dominated by physical escape and adiabatic losses. The second flare is also less bright than the first one, which is another manifestation of this effect.

Radio flares have very different behavior in these reaccelerating episodes. Contrary to the $\gamma$-ray flares, which appear with various shapes and efficiencies, radio frequencies might show no flaring activity at all and this depends critically on the distance where the episode takes place. According to our analysis presented in Section \ref{s3}, a reaccelerating episode will appear in radio only if it occurs beyond the distance where the emission becomes optically thin to synchrotron self-absorption. This is exemplified in Fig.\ref{fig:thickness} which shows two multiwavelength flares, one induced by electrons injected in an optically thick region (Fig.\ref{fig:thicknessa}) and another in an optically thin (Fig. \ref{fig:thicknessb}). We plot the flaring lightcurves (full line) ranging from radio to $\gamma$-ray in both figures. For comparison, we also plot the lightcurve we would get without the reaccelerating episode. One could see that when the episode occurs closer to the origin (Fig. \ref{fig:thicknessa}), a flare is detected in $\gamma$-ray, X-rays and optical but not in radio frequencies. However, when the episode occurs beyond the transition radius, then a flare appears at all frequencies (Fig. \ref{fig:thicknessb}). Note that in radio higher frequencies appear first.  

\subsection{Application to the flare of Mrk421}
Mrk421 is one of the most well-studied blazars. In April 2013, the source was in intense activity recorded from X-rays to TeV energies. Following this detection the Combined Array for Research in Millimeter-Wave Astronomy (CARMA) observations started. According to \cite{HP15} the activity of the source in the $\gamma$-ray started in March 2013, while the lightcurve reached its peak on April 14, 2013. The photon flux almost quadrupled within 30 days, while the total activity of the event was about 100 days. In radio frequencies, the intense activity started in April and reached its maximum flux in June with an increase of about a factor of 1.5. In radio frequencies, the source was recorded at 15 GHz and 95 GHz with a slight lag of the two peaks and a complete correlation between them. However, in the data analysis, it appeared that there was also a correlation between $\gamma$-ray\footnote{Data by \emph{Fermi} $\gamma$-ray telescope.} and radio.

\begin{figure}[!htbp]
%	\addtocounter{figure}{-1}
\centering
\begin{subfigure}[b]{0.45\textwidth}
\centering
\includegraphics[width=9cm, trim=60 4 4 6]{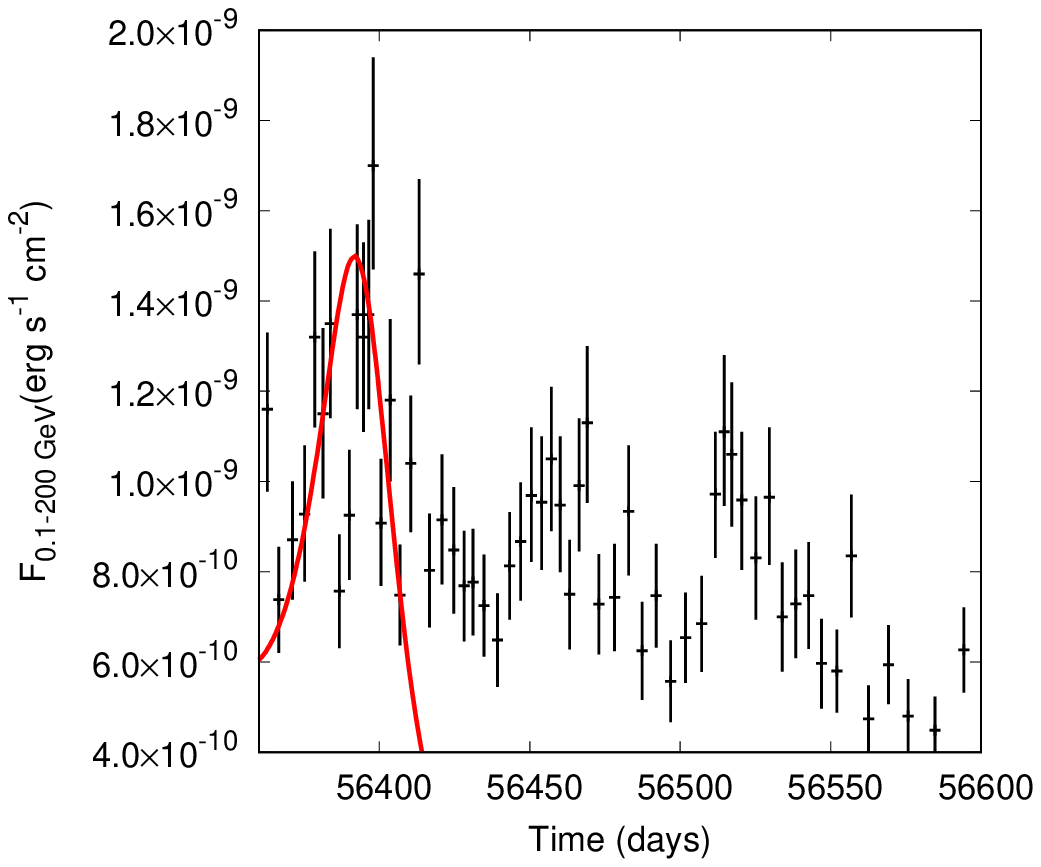}
\caption{}\label{fig:mrk421g}
\end{subfigure}
\hspace{1cm}
\begin{subfigure}[b]{0.45\textwidth}
\centering
\includegraphics[width=9cm, trim=60 4 4 6]{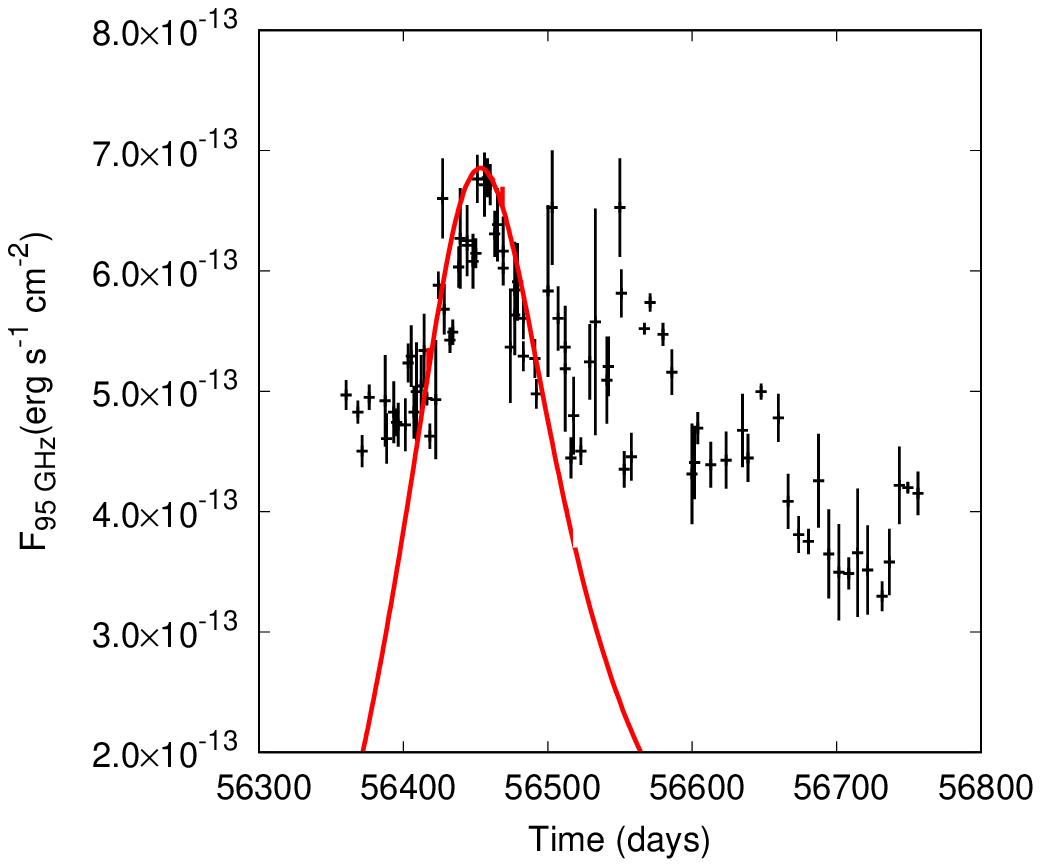}
\caption{}\label{fig:mrk42195}
\end{subfigure}
\caption{Modeling of Mrk 421 2013 flare. Observational data are presented at \cite{HP15}. The parameter set that we use in this modeling are:
$R'_{\rm 0}=10^{16.6}$ cm, $B'_{\rm 0}=1.25$ G, $L_{\rm e_{\rm 0}}^{\rm 'inj}=10^{43.2}$ erg s$^{-1}$, $t'_{\rm esc,e}=R'/c$, $u_{\rm exp}=0.05$ c, $\gamma_{\rm min}=1$, $\gamma_{\rm max}=10^{5.1}$, $p=2$, $\delta=10$ and $z_{\rm 0}=0.1$ pc. Magnetic field decreases as $B'\propto R'^{-1}$ and electrons luminosity as $L_{\rm e_{\rm 0}}^{\rm 'inj}\propto R'^{-2}$. For the $\gamma$-ray flare we used a pulse which was injected at $t'_{*}=40~t_{\rm cross,0}$, with a width $w'=23~t_{\rm cross,0}$ and $\alpha=60$.}%
\label{fig:Mrk421flare}%
\end{figure}

Motivated by this analysis, we model the observational data as recorded in the 2013 Mrk 421 flaring activity. Our approach to studying this particular flare is by applying the method described earlier. An energetic plasma blob is produced at a distance $z_0=0.1$ pc from the origin and moves along its axis. The $\gamma$-ray (Fig. \ref{fig:mrk421g}) are produced by a particle reaccelerating episode at $z_*=0.6$ pc through the SSC process. At this distance, the jet is opaque to radio emission due to the synchrotron self-absorption. However, as the plasma blob expands, it becomes optically thin and the radio waves at 95 GHz escape at $z_{ssa}(95 ~GHz)\simeq 1.1$ pc from the origin and this transition is recorded as a flare  at the specific frequency (see Fig. \ref{fig:mrk42195}). The total electron energy required to power this flare is $\sim 10^{49}~{\rm erg}$. More blobs could explain the rest activity of Mrk421. Despite the fact that the fit is not perfect, it should be compared to a similar fit presented in Fig.4 of \cite{HP15}, which was obtained using the non-expanding version, \cite{MK95}, of the present code.

\section{Discussion}\label{s5}

This paper presents an extension of the one-zone leptonic model by including expansion of the source. 
We have based our numerical scheme on the time-dependent model of \cite{MK95} which uses a spherical geometry and
has the time and energies of electrons/photons as independent parameters. The present code can describe expansion by replacing the time coordinate with the source's radius assuming a constant expansion velocity. Therefore it solves the kinetic equations that describe the evolution of the electron and photon spectra at each source radius, or, equivalently, at each distance of the source from the origin.

The kinetic equations are the same as in \cite{MK95} - with the additions/refinements made in subsequent publications, and describe the usual processes of a one-zone leptonic model, i.e., synchrotron and inverse Compton both as energy losses for electrons and sources for photons plus photon-photon interactions both as an injection term for electrons and as a sink term for photons. In addition we have used synchrotron self-absorption for photons and a physical escape term for electrons. Furthermore, the expansion allows us to take explicitly adiabatic losses, which in many non-expanding codes are taken simply as an escape term over the dynamical timescale.

The code, by construction, does not provide a steady-state solution but rather a series of snapshots of the photon spectrum of the source (blob) as it travels down the jet. It also requires more free parameters than the non-expanding model.
Thus, one needs prescriptions for the profile of the total power injected in electrons and for the magnetic field strength with distance from the origin and a value for the expansion velocity of the blob. These come 
in addition to the usual parameters of the leptonic model which include the radius of the blob, its magnetic field strength and the details of the electron injection distribution (total injected power, upper and lower energy cutoffs and slope in the case of a power-law distribution). 
Despite the fact that there are ways that one could relate at least some of these parameters physically \citep[see, e.g.,][]{Boula19b}, for the purpose of the present paper, we have used a more generic approach and we have left them as free. Note that we do not vary with the distance the shape of the electron distribution (high and low energy cutoffs plus power-law slope) because this would have added more free parameters or, alternatively, it would have required a detailed acceleration scheme that we have not included in the present version of the code.

As a first example, we apply the expanding code to treat the radio emission of blazars. This is usually problematic for the non-expanding one-zone models since they provide fits that require a small size for the source and thus a high electron number density. This leads, in turn, to an intense synchrotron self-absorption of the radio photons and one requires a different region to produce them \citep[e.g.,][]{M80,G85,MG85}. With the present code we show that while high-energy photons can be produced close to the base of the jet where the bulk of the injection power takes place, radio photons of some frequency appear when the radius of the blob is such that the optical depth at that particular frequency drops below unity. This occurs at much larger distances from the origin where the blob has opened up and both the electron density and the magnetic field have dropped. We find that this distance depends on many parameters, but the rule of thumb is that it appears further away from the origin for higher initial values of the electron luminosity and magnetic field strength. If therefore, both of these quantities drop outwards, one expects high-energy radiation to be produced very close to the site of the blob formation, while radio photons appear much further away.

We should also emphasize that adiabatic losses can become dominant as the blob expands, and this has severe consequences on the radiative efficiency of the electrons -- at large distances from the origin, even freshly injected electrons become inefficient emitters. This can be seen, mostly at higher frequencies, even during the duration of a flare. Generally speaking, we find that more asymmetric flares imply high expansion velocities as parameters like the magnetic field can vary a lot during an episode of electron injection. Therefore it is conceivable that, e.g., a $\gamma$-ray\footnote{All $\gamma$-ray shown in the paper are produced from the SSC mechanism, therefore the examples refer to BL Lac objects. FSRQs, on the other hand, are expected to have external photon fields, which is expected to affect the time behavior of the $\gamma$-ray.} flare starts when electrons are dominated by radiative losses and ends when they are in the adiabatic phase\footnote{Electron physical escape might also play a role -- see the analysis presented in\cite{2002Wang}.} . In the non-expanding case, such asymmetries can be achieved only by fine-tuning the electron escape timescale from the source. However, we should mention that we have not considered the light travel time effects between photons emitted from the back and the front side of the blob, which are also expected to play a role in the flare shape, \cite{2013Zacharias}.
 
 As a second example, we have added an injection of relativistic electrons in the form of a Lorentzian, which occurs on top of the power-law decay of electron power that we used in the previous example. We use this approach in order to study the relation of the produced $\gamma$-ray and radio flares. Even this simple case shows clearly that one can detect both $\gamma$-ray and radio in the case when the reaccelerating episode occurs in a region that is optically thin to radio frequencies. If the episode occurs further in, only $\gamma$-ray (and other high-frequencies) flares are detected but not radio flares \footnote{On the other hand, if we suppose that the high-frequencies peak luminosities are below the steady-state emission, it is possible only the radio to appear above the steady-state and be detected as a flare.}. While it is possible for one to make a parametric study, here, we have refrained from doing so due to the large number of parameters involved, which would have obscured our results. However, as we have shown in \cite{Boula19b} one expects a wide range of time lags when both $\gamma$-ray and radio flares appear. As a final example, we have shown a fit to a particular radio and $\gamma$-ray flare of Mkr 421 that occurred in 2013, \cite{HP15}.
 
 While this code is best suited to follow the radiative patterns of accelerating episodes as they move down the jet, it does not provide the SED for a global jet. One could envisage a steady-state jet spectrum by a superposition of the spectra of blobs created at a constant rate with the same initial conditions that emit continuously as they travel along the jet \cite[][]{Boula19b} but here we do not deal with this option.
One should mention that there are other numerical approaches to calculate the produced photons in the case of blazar jets \citep[e.g.,][]{potter1,2011tramacere,2021zacharias}. Specifically, \cite{Potter18} have presented similar results using a different geometry, however for the emitting region. 

Concluding we can say that the expanding one-zone leptonic model presented in this paper is inherently different from the usual non-expanding one. While the latter has fewer free parameters and can reach a steady-state, the former provides perhaps a more realistic prescription for the blob evolution as it takes into account expansion, fully-fledged adiabatic losses and a varying source's light-crossing time. However, this happens at the expense of more free parameters that could partly obscure the physical mechanisms at the source -- this is particularly true in the case where a flaring episode is simulated. In the present paper we have shown only a few characteristic examples emphasizing the connection between radio and $\gamma$-ray frequencies and we have drawn some general conclusions. We believe that other conclusions can be drawn only by fitting data on a source-to-source basis using both temporal and spectral information.

\section{Conclusions}\label{s6}
In this paper, we present a new numerical code that describes a one-zone leptonic expanding model and we use it to investigate some general trends of the multiwavelength activity of blazars. More precisely we assume a spherical blob of plasma that expands and propagates down along the blazar jet's axis. We solve self-consistently the time-dependent integro-differential kinetic equations that describe the evolution of particles and photons by taking the leptonic physical processes into account and by assuming that all physical quantities are related to the co-moving time. 
By applying this code to blazar jets we are able to:
\begin{itemize}
\item Localize the distance of radio emission onset depending on the fundamental physical quantities of the source, such as the magnetic field, the electron luminosity and the initial source's radius at the base of the jet.
\item Connect radio and $\gamma$-ray flares by assuming electron accelerating episodes. 
\end{itemize}
This numerical tool can be used for a comprehensive study of flaring activity, e.g., in the case of blazars, both for BL Lac objects and Flat Spectrum Radio Quasars (where we consider external photon fields). Also, it can be applied to compact jetted sources, such as active galactic nuclei in general, micro-quasars, tidal disruption events and gamma-ray bursts.

\begin{acknowledgements}
SB: This research is co-financed by Greece and the European Union (European Social Fund-ESF) through the Operational Programme "Human Resources Development, Education and Lifelong Learning" in the context of the project "Strengthening Human Resources Research Potential via Doctorate Research" (MIS-5000432), implemented by the State Scholarships
Foundation (IKY). We thank the anonymous referee for the useful comments. We thank Dr. Maria Petropoulou for her helpful comments on the manuscript and discussions. Also, we thank Dr. Tavliki Hovatta, who provides us the observational data of Mrk421.  
\end{acknowledgements}

\bibliographystyle{aa} % style aa.bst
\bibliography{bib} % your references Yourfile.bib
\end{document}